\documentclass[preprint]{aastex6}
\newcommand{\mmsm}{mm/submm-$\lambda$}

\newcommand{\mgh}{Mg II h line\ }
\begin{document}

\title{A First Comparison of Millimeter Continuum and M\MakeLowercase{g} II Ultraviolet Line Emission from the Solar Chromosphere}

\author{T. S. Bastian\altaffilmark{1}, G. Chintzoglou\altaffilmark{2}, B. De Pontieu\altaffilmark{2,3}, M. Shimojo\altaffilmark{4,5}, D. Schmit\altaffilmark{2,6}, \\ J. Leenaarts\altaffilmark{7}, M. Loukitcheva\altaffilmark{8,9}
}
\affil{$^1$National Radio Astronomy Observatory, 520 Edgemont Road, Charlottesville, VA 22903, USA \\ 
$^2$Lockheed Martin Solar \& Astrophysics Lab, Org. A021S, Bldg. 252, 3251 Hanover Street, Palo Alto, CA 94304, USA\\
$^3$Institute of Theoretical Astrophysics, University of Oslo, Blindern, Blindern, P.O. Box 1029 Blindern, N-0315 Oslo, Norway\\
$^4$National Astronomical Observatory of Japan, 2-21-1, Osawa, Mitaka, Tokyo 181-8588, Japan\\
$^5$Department of Astronomical Science, The Graduate University for Advanced Studies (SOKENDAI), 2-21-1, \\ Osawa, Mitaka, Tokyo 181-8588, Japan\\
$^6$Bay Area Environmental Research Institute, 625 2nd St. Ste 209 Petaluma, CA 94952 USA\\
$^7$Institute for Solar Physics, Department of Astronomy, Stockholm University, AlbaNova University Centre, SE-106 91 Stockholm, Sweden\\
$^8$Center For Solar-Terrestrial Research, New Jersey Institute of Technology, 323 Martin Luther King Boulevard, Newark, NJ 07102, USA \\
$^9$Max-Planck-Institut f\"ur Sonnensystemforschung, Justus-von-Liebig-Weg 3, 37077, G\"ottingen, Germany \\
}
\email{tbastian@nrao.edu}
\shorttitle{Millimeter and UV Emission from the Chromosphere}
\shortauthors{Bastian et al.}

 
\begin{abstract}

We present joint observations of the Sun by the Atacama Large Millimeter/submillimeter Array (ALMA) and the Interface Region Imaging Spectrograph (IRIS). The observations were made of a solar active region on 2015 December 18 as part of the ALMA science verification effort. A map of the Sun's continuum emission of size $2.4' \times 2.3'$ was obtained by ALMA at a wavelength of 1.25 mm (239 GHz) using mosaicing techniques. A contemporaneous map of size $1.9'\times 2.9'$ was obtained in the Mg II h doublet line at 2803.5\AA\ by IRIS. Both \mmsm\ continuum emission and ultraviolet (UV) line emission are believed to originate from the solar chromosphere and both have the potential to serve as powerful and complementary diagnostics of physical conditions in this poorly understood layer of the solar atmosphere.  While a  clear correlation between \mmsm\ brightness temperature $T_B$ and the \mgh radiation temperature $T_{rad}$ is observed the slope is $<1$, perhaps as a result of the fact that these diagnostics are sensitive to different parts of the chromosphere and/or the Mg II h line source function includes a scattering component. There is a significant offset between the mean $T_B$(1.25~mm) and mean $T_{rad}$(Mg II), the former being $\approx 35\%$ greater than the latter.  Partitioning the maps into ``sunspot", ``quiet regions", and ``plage regions" we find that the slope of the scatter plots between the IRIS \mgh $T_{rad}$ and the ALMA brightness temperature $T_B$ is 0.4 (sunspot), 0.56 (quiet regions), and 0.66 (plage regions). We suggest that this change may be caused by the regional dependence of the formation heights of the IRIS and ALMA diagnostics, and/or the increased degree of coupling between the UV source function and the local gas temperature in the hotter, denser gas in plage regions.

\end{abstract}



\section{Introduction} \label{sec:intro}

The solar chromosphere is a poorly understood region of the solar atmosphere in which non-radiative heating becomes manifest. It is a highly dynamic layer through which acoustic waves propagate from below and shock, dissipating their energy. The chromosphere is also permeated by complex magnetic fields. The plasma $\beta$ parameter, the ratio of gas pressure to magnetic pressure, varies by orders of magnitude within the chromosphere, as does the ionization fraction. Hence, establishing the structure and dynamics of the solar chromosphere and understanding the transfer of radiative and mechanical energy through the chromosphere pose significant challenges. 

Optical and UV line emission, as well as infrared, millimeter, and submillimeter wavelength continuum emission, originate from the chromosphere and have the potential to serve as powerful probes of its structure and dynamics but their utility has sometimes been limited by either their accessibility to observation (UV)  or by technical challenges (\mmsm).  Recently, high resolution observations of the solar chromosphere in UV lines became accessible thanks to the Interface Region Imaging Spectrograph (IRIS) small explorer space mission (De Pontieu et al. 2014). Among the newly available UV lines, the Mg II h and k doublet lines are of particular interest. Magnesium is more abundant than calcium in the solar atmosphere and it therefore samples a wider range of heights of the solar chromosphere than other common line diagnostics such as the Ca II H and K lines, the He I 10830\AA\ line, or H$\alpha$ (Leenaarts et al. 2013a). Interest in the Mg II h and k lines as a means of diagnosing key parameters at chromospheric heights has therefore been strong and effort has been focused on modeling their radiative transfer in the context of chromospheric models  (Leenaarts et al. 2013ab; Pereira et al. 2013; Sukhorukov \& Leenaarts 2017). 

Continuum observations at millimeter and submillimeter wavelengths (mm/submm) are also an attractive probe of the chromosphere (e.g., Bastian 2002). The dominant sources of opacity -- collisions between free electrons and ions (free-free opacity) and between free electrons and neutral hydrogen (H$^-$ opacity) -- are well understood; the radiation is formed under conditions of LTE and the source function is therefore the Planckian. Since $h\nu/KT\ll 1$ the Rayleigh-Jeans approximation is valid and the observed intensity is linearly proportional to the temperature of the emitting material. We note that the degree of ionization for hydrogen and helium is not in statistical equilibrium in the solar chromosphere (Carlsson \& Stein 2002; Leenaarts et al. 2007; Golding et al. 2016); while the source function is in LTE, the opacity can be far from its LTE value (Wedemeyer-B\"ohm et al. 2007).The primary disadvantage to observations at \mmsm has been the relatively poor angular resolution available in past years as a result of the fact that most observations were acquired using single dishes (e.g., White et al. 2017). 

With the advent of the the Atacama Large Millimeter/submillimeter Array (ALMA), high resolution observations of the Sun are now possible and direct comparisons with other diagnostics of the solar atmosphere have become feasible. In this Letter we present a first comparison of solar observations of the chromosphere in a solar active region made by ALMA in the 1.25 mm band (239 GHz) with those of the same region made in the ultraviolet Mg II h doublet line by  IRIS. In \S2 we briefly describe the instruments and the observations. In \S3 we present maps of the 1.25 mm continuum brightness temperature and the UV Mg II h2v radiation temperature. We discuss mm/UV-wavelength correlations for various regions and suggest underlying reasons for their attributes. We conclude in \S4.

\section{Observations and Analysis} \label{sec:obs}

\subsection{Atacama Large Millimeter/submillimeter Array}\label{sec:ALMA}

ALMA is a powerful, general purpose interferometer comprised of the 12m array (50 $\times 12$m antennas) and the Atacama Compact Array (ACA: $12 \times 7$m antennas and $4 \times 12$m total power antennas).  ALMA is designed to perform imaging  observations and spectroscopy of astrophysical phenomena in frequency bands that will ultimately span 35-950 GHz, or wavelengths of 0.32-8.6 mm. Limited solar observing modes were released to the scientific community in 2016 (ALMA Cycle 4). Specifically, continuum observations are possible in band 3 (100 GHz/3~mm) and band 6 (239 GHz/1.25~mm). In order to observe the Sun with ALMA, special hardware modifications and calibration procedures were developed. The testing and commissioning effort that enabled solar observing with ALMA in bands 3 and 6 are described in detail by Shimojo et al. (2017) and White et al. (2017).  

The final solar testing campaign prior to the release of solar observing modes in ALMA Cycle 4 was conducted in December 2015. A variety of targets and modes were observed to ensure that systems performed as expected. Among the observations were band 6 observations of NOAA active region number 12470 on 18 December 2015 from 19:11:41 - 20:09:59 UT when the active region was at approximately N15E05. The band 6 observations were acquired in 4 spectral windows, each 2 GHz in bandwidth, centered on 230, 232, 246, and 248 GHz. The mean frequency was 239 GHz, or a wavelength of 1.25~mm. The instantaneous field of view (FOV) of ALMA is determined by the primary beam of a single antenna: for band 6 the full width at half maximum of an ALMA 12m antenna is $24"$. While this is sufficient for some types of observing programs, the restricted FOV hampers others. For programs that require mapping an area larger than the antenna FOV, mosaicing techniques are used where the interferometer points to a grid of locations that are specified relative to a reference pointing. In doing so, care must be taken to correct the individual mosaic pointings for the physical ephemeris of the Sun and the Sun's differential rotation. The ALMA band 6 map of AR~12470 presented here was formed from a mosaic of 149 pointings (Nyquist sampled), each pointing of duration 6.048s. 

In order to produce maps calibrated in absolute terms, units of flux density or brightness temperature, it is necessary to recover all angular scales present in the target. An interferometer acts as a high-pass spatial filter. A map produced by interferometry alone filters out angular scales larger than those measured by the minimum  spacing between two antennas. ALMA was carefully designed to recover all relevant angular scales in most cases, but for sources that are complex and extended it is necessary to use both interferometric and total power (TP) measurements. The TP antennas enable maps of the Sun on the largest angular scales, from the angular resolution of a single 12m antenna to the angular scale of the Sun itself ($\approx 1920"$), allowing power on all spatial frequencies to be  restored. Calibration of the TP measurements is described by White et al. (2017) where the brightness temperature of the quiet Sun at the center of the solar disk is scaled to 5900 K (1.25~mm). The interferometric and TP mapping data were combined using ``feathering" techniques as described by Shimojo et al. (2017).  The 1.25 mm map of AR 12470 produced on 18 December 2015 is $143" \times 139"$ in size and has an angular resolution of $2.4"\times 0.9"$ (position angle -76.7$^\circ$). 
The photometric calibration of the resulting map is dominated by the total power mapping measurements. These are expected to be calibrated in absolute terms to $\approx 5\%$ (see White et al. 2017). 

\subsection{Interface Region Imaging Spectrograph}\label{sec:IRIS}

The Interface Region Imaging Spectrograph (IRIS; De Pontieu et al 2014) conducted high spatial ($0.16"$ pix$^{-1}$), temporal and spectral resolution (0.025\AA) observations of the Mg II h line at 2803.5\AA\ of AR 12470 in coordination with ALMA.  The IRIS data presented here were obtained on 2015 December 18 starting at 19:33:13 UT. They are comprised of three dense ($0.35"$) 320-step very large raster scans (FOV $112" \times 175"$) centered at $(x,y)=(-59" \times 231")$. Spatial binning ($\times 2$) was performed on the slit dimension and also on the spectra (again $\times 2$). The exposure time was 2 s and each raster scan was obtained in 1,026 s (with a total duration for the observation of 3078 s). We present the first large, dense raster -- from 19:33:13 to 19:51:09 UT.  All spectral data sets have been organized as level-2 FITS files and were already corrected as appropriate for the instrumental effects (such as dark subtraction, flat-fielding, geometric and wavelength corrections). 

We limit our analysis here to the Mg II h doublet line at 2803.5\AA.
The average solar Mg II h line profile is an emission line with a reversed core (yielding red ``h2r'' and a blue ``h2v'' peaks on either side of the ``h3'' core). However, pertinent to the local conditions, the opacity modulates the double-peaked profile, making it asymmetric or even single peaked. We used a nine-parameter double-Gaussian model (Schmit et al. 2015) based on MPFIT (Markwardt 2009) to model the line profile by a linear combination of a wide positive-amplitude Gaussian and a negative narrower-width Gaussian. We stored the model fits in separate arrays for the intensities and wavelengths for both the red and blue Mg II h2 peaks. Wherever the line core was not depressed, thus rendering the Mg II h line a single-peaked profile, the intensity and wavelength of that single peak was recorded in the arrays (in a way to prevent missing values in our rasters).

We then used the model fits for the intensity rasters and performed radiometric calibration to convert the intensity from DN s$^{-1}$ to physical intensity units (in erg s$^{-1}$ cm$^{-2}$ sr$^{-1}$ \AA$^{-1}$). We used the latest calibration values to date for this purpose (i.e. version ``004'').  Using the radiometrically calibrated intensity rasters -- absolute calibration of order 15\% -- we derived the radiation temperatures $T_{rad}$ for each peak that we used in our analysis.

\begin{figure}[ht!]
\begin{center}
\figurenum{1}
\includegraphics[width=6.5in]{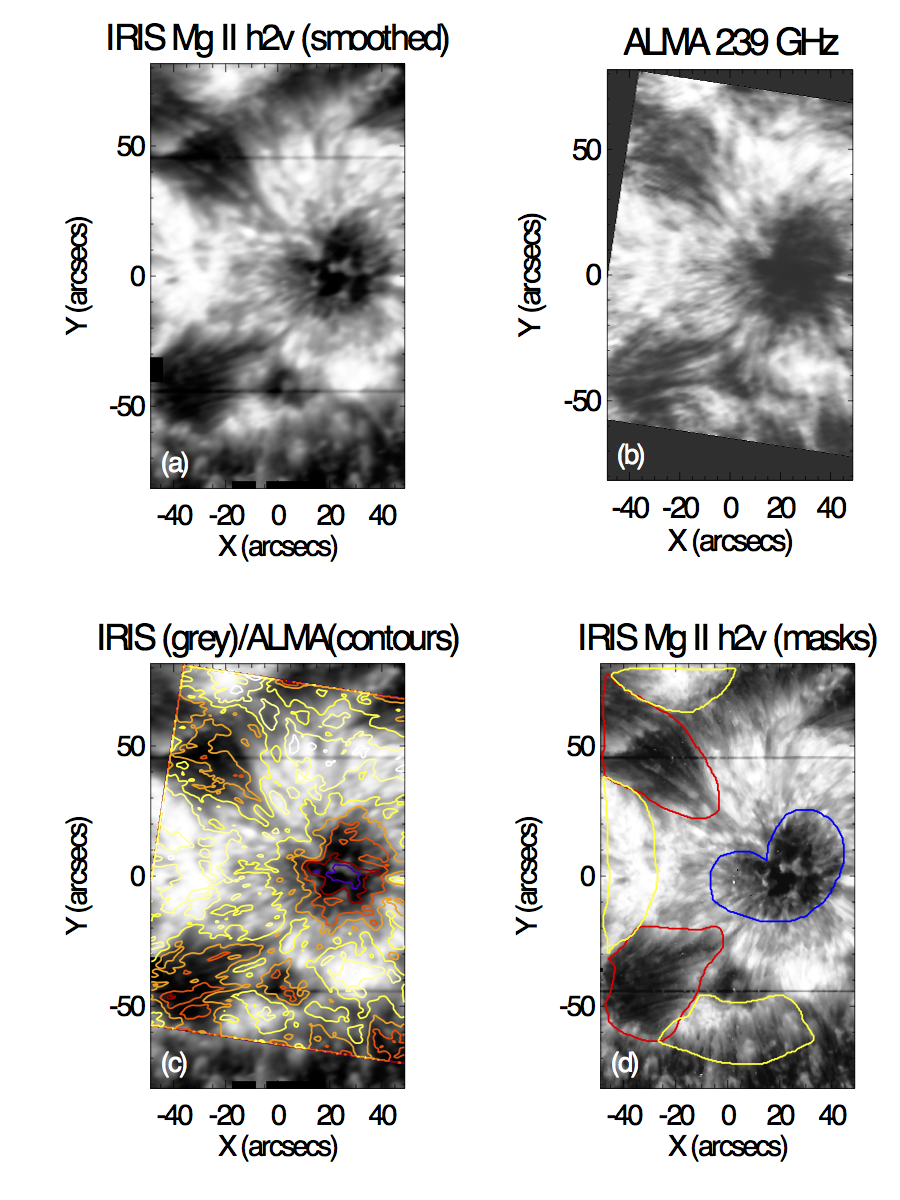}
\caption{(a) A map of the IRIS Mg II h2v radiation temperature convolved to the resolution of the ALMA map; (b) a map of the the ALMA 1.25 mm continuum brightness temperature; (c) a comparison between the IRIS map of $T_{rad}$ (greyscale) and the ALMA map of $T_B$ (contours); (d) A map of the IRIS Mg II h2v line radiation temperature showing the masked areas corresponding to the sunspot (blue), quiet regions (red), and plage regions (yellow).}
\end{center}
\end{figure}

\bigskip\bigskip

\section{Results and Discussion} \label{sec:disc}

In order to compare the IRIS raster data with ALMA imaging data several steps were taken. The ALMA map was obtained in the geocentric coordinate frame (right ascension and declination). It was therefore necessary to correct the final map for the solar position angle. For IRIS,  pixels for which the fitting procedure failed ($<0.1\%$) were filled through the use of a median filter. There was a small amount of position ``offset" from raster to raster.
To correct for this, the IRIS raster data were interpolated onto a uniform grid ($0.4"\times 0.4"$ pixels). The resulting map of $T_{rad}$  in the Mg II h2v line was then convolved with a Gaussian kernel so that it had the same angular resolution as the corresponding ALMA map. We note that the conversion of the Mg II h2v line intensity to radiation temperature removes the nonlinearity of the Planck function in UV so that the subsequent smoothing step does not bias the radiation temperature, as would have been the case had the steps had been reversed. Finally, the maps were corrected for position offsets and differing FOVs to bring them into alignment to allow comparisons to be made on a pixel by pixel basis. Note, however, that we have not corrected for the fact that pixels are correlated on the scale of ALMA's resolution; i.e., the pixel size is $0.4"\times 0.4"$ and they are correlated on an angular scale of $2.4"\times 0.9"$. 
 
The matched-resolution maps obtained by ALMA and IRIS are shown in Fig.~1. A clear correlation is evident between the ALMA map of the 1.25~mm continuum brightness temperature $T_B$ and the IRIS map of the Mg II h2v radiation temperature $T_{rad}$. To better characterize the nature of the correlation we formed scatter plots of the ALMA and IRIS observations as shown in Fig.~2. The central panel shows a scatter plot of all pixels in common to the two maps (grey pixels). While the trend remains clear, the scatter of pixels is significant. The Pearson correlation coefficient for all pixels is 0.78. We note that the mean and range of 1.25~mm continuum brightness temperatures are approximately 7380 K and 5200-8700 K, respectively,  whereas the mean and range of the Mg II h2v radiation temperatures are approximately 5470 K and 4500-6500 K, respectively. Hence, there is a significant offset between  $T_B(\rm 1.25~mm)$ and $T_{rad}$(Mg II), and a compression in range of the latter relative to the former.  A linear fit to the data yields a slope of $m=0.57$. The causes for the slope, the offset and the degree of scatter between the 1.25~mm brightness temperature and the radiation temperature derived from Mg II h2v are not fully clear. Preliminary results from advanced numerical models (e.g., those computed with the Bifrost model: Gudiksen et al. 2011; see also Carlsson et al. 2016 and references therein) of quiet Sun conditions indicate that the Mg II h2v feature forms in the middle chromosphere over a range of heights from 600 to 2,000 km with a peak around 1,400 km (Leenaarts et al., 2013b). Synthetic observables calculated from the Bifrost at a wavelength of 1.3~mm, near that reported here, suggest that the radiation mostly originates at similar heights, with a peak at 1150 km (Loukitcheva et al., 2017). While the diagnostics are formed in roughly the same region of the atmosphere, it is unclear whether there are systematic differences in formation height along individual lines of sight. For both diagnostics the formation height surface is highly corrugated with large variability expected from location to location. It is possible that the height difference between these diagnostics is larger than predicted by the models. There are already indications that the Bifrost model does not properly reproduce the middle to upper chromosphere where significant opacity for both IRIS and ALMA diagnostics occurs, as exemplified by the large discrepancy in line width between observed and synthetic Mg II h profiles from the simulations (Leenaarts et al. 2013a). In particular, the Bifrost model appears to lack opacity from upper chromospheric features such as spicules and fibrils. Hence, one possibility is that the observed slope is in part caused by a differing sensitivity of the diagnostics to local conditions in such features, with ALMA more sensitive and IRIS less sensitive because of the scattering nature of the Mg II line. It is also possible that the compressed temperature range deduced from the optically thick Mg II h2v feature could be caused by radiative transfer effects. Since the Mg II h and k lines are scattering lines, the source function has a scattering contribution that leads to a source function that is not coupled to the local conditions and is lower than the local Planck function. Leenaarts et al. (2013b) also show that while the model correlation is good between the Mg II line radiation temperature and gas temperature, the radiation temperature underestimates the gas temperature by $\sim500$ K (much smaller than the offset between $T_{rad}$(Mg II) and $T_B$(1.25~mm) observed). However, at low intensities the correlation shows a significant scatter with many pixels having gas temperatures that are significantly higher than $T_{rad}$(Mg~II) -- again the result of radiation scattering. While the slope and scatter of the correlation may be qualitatively understood, in part, the significant offset between $T_{rad}$(Mg II) and $T_B(\rm 1.25~mm)$ is not understood within the context of current modeling efforts. We caution that it is in any case unclear how applicable these quiet Sun models are to our observations of AR~12470, where even the quietest regions are affected by the super-penumbra that surrounds the sunspot and plage. Nevertheless, taken together these observations thus provide stringent constraints on future numerical models of the chromosphere.

It is interesting to compare various regions in the ALMA and IRIS maps. To do so, areas of AR 12470 mapped by both instruments were identified as corresponding to the sunspot, ``quiet'' regions (see comment above about the impact of the super-penumbra), and plage regions. Fig.~2d shows the masked areas corresponding to the sunspot (blue contour), to ``quiet'' regions (red contour), and to plage regions (yellow contour). 
Considering each of these regions separately shows that the correlation between the 1.25~mm brightness temperature and the Mg II h2v line radiation temperature is nonlinear. The solid line in Fig.~2 shows a linear fit to sunspot pixels (slope $m=0.4$); the dashed line shows the same for ``quiet'' regions pixels ($m=0.56$), and the dot-dash line shows the same for plage regions ($m=0.66$). The reason for this behavior is not fully understood. In an analysis of IRIS Mg II k line profiles in plage regions Carlsson et al. (2015) suggest that the observed line profiles arise naturally in hot, dense plasma, ensuring better coupling of the source function to the local temperature. Hence, the increased slope observed in the correlation between the 1.25~mm continuum emission and the Mg II h2v line may similarly reflect a higher degree of coupling of the source function to the local gas temperature. 

We have checked sources of possible systematic error in both datasets to ensure that the attributes we have described are robust. Chief among these is the absolution calibration of the two datasets. As noted in previous sections, the ALMA brightness scale is dominated by the total power measurements, believed to be accurate to $\approx 5\%$. Radiometric calibration of the IRIS observations is of order 15\%. Another possible source of error is the fact that the two data sets not obtained strictly simultaneously - the IRIS data were obtained during the course of $\sim18$ min during ALMA mosaic mapping ($\sim 50$ min duration). While time variability likely increases the scatter in the correlation between the two datasets, it cannot account for the offsets and slopes of the scatter plots. In the case of IRIS, using contemporaneous sparse raster data, we estimate an rms of $\sigma=300$ K in the sunspot umbral region (largely due to oscillations) and $\sigma=200$ K in plage regions. ALMA data at 1.25~mm suitable for characterizing the time variability of the continuum mm emission are not available for this active region. A study of the mm and UV time variability in sunspot, quiet, and plage regions will be pursued separately. 
Another consideration In the case of ALMA is whether the contribution of the overlying corona to the observed brightness temperature, $\Delta T_B$(1.25~mm), is significant. The dominant source of opacity over the active region at mm-$\lambda$ is free-free absorption. The optical depth of a differential layer $dz$ overlying the active region is $d\tau\approx 0.2 (\lambda/c)^2 n_e^2 T^{-3/2} dz$. The brightness temperature contribution of the overlying (optically thin) coronal material at 1.25~mm was then estimated as 
 
 \begin{equation}
 \Delta T_B (1.25~{\rm mm}) \approx \int T d\tau \approx 0.2{\lambda^2\over c^2} \int T^{-1/2} n_e^2 dz = 0.2{\lambda^2\over c^2} \int T^{-1/2} \phi(T) dT
 \end{equation}
 \medskip
 
 \noindent where $\phi(T)=n_e^2(T) dz/dT$ is the differential emission measure of the coronal material over AR 12470. We estimated $\phi(T)$ using measurements by the Solar Dynamics Observatory  Atmospheric Imaging Assembly and found that $\Delta T_B$(1.25~mm) was no more than 2\%. We conclude that none of the potential sources of systematic error likely has a significant impact on the ALMA/IRIS correlation properties observed. 
 
\begin{figure}[ht!]
\begin{center}
\figurenum{2}
\includegraphics[width=6.5in]{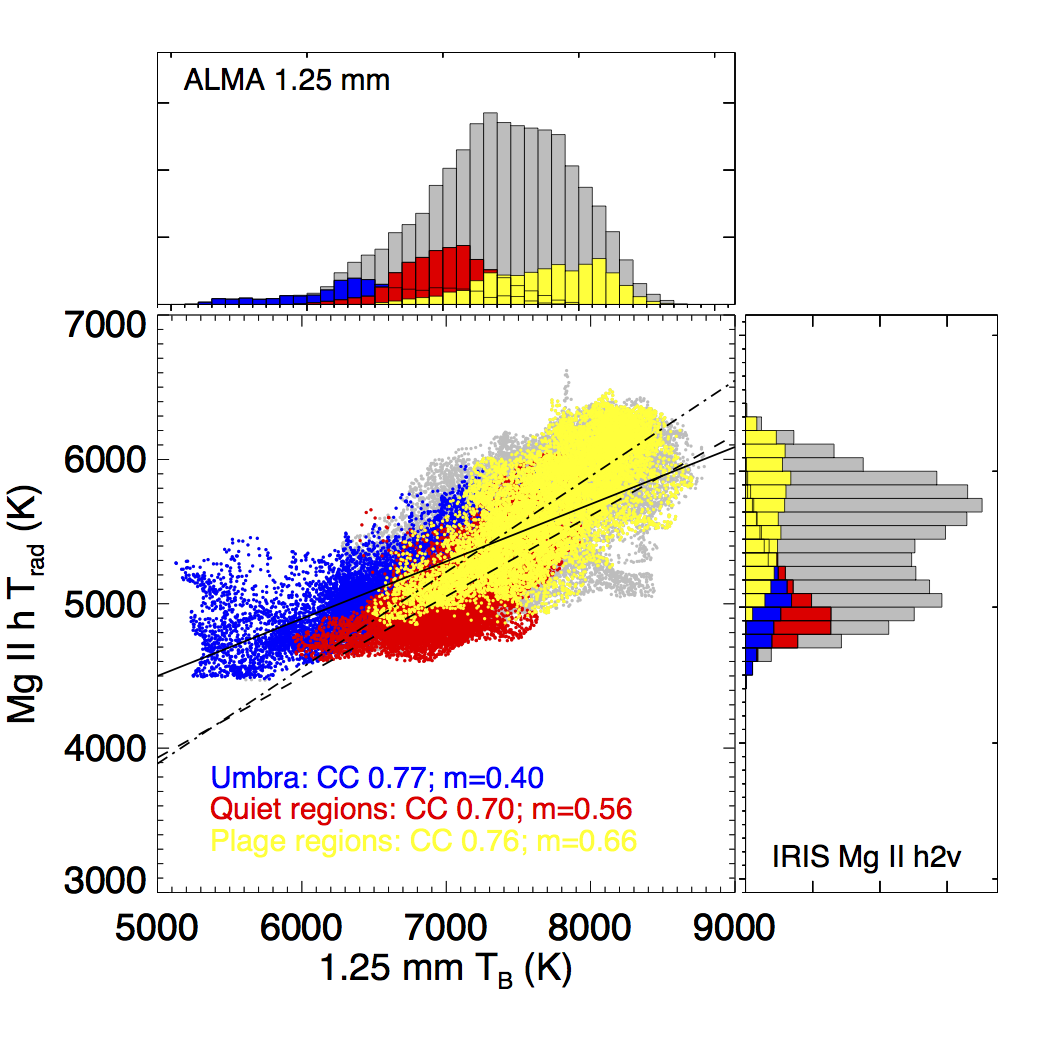}
\caption{The main panel shows a scatter plot of IRIS Mg II h2v radiation temperatures versus the corresponding ALMA 1.25 mm continuum brightness temperatures. The grey points represent all common pixels in the two maps. The blue points represent pixels in the sunspot umbra; red points represent pixels in ``quiet'' Sun areas; and yellow pixels represent pixels in plage regions. The corresponding histograms are shown for the ALMA data (top panel) and the IRIS data (right panel). The mean 1.25~mm brightness temperature of the sunspot, quiet, and plages regions are 6555~K, 7045~K, and 7735~K, whereas the mean Mg II h line radiation temperatures are 5116~K, 5076~K, and 5707~K, respectively. Linear fits to pixels representing the sunspot umbra, ``quiet'' regions, and plage regions are shown as solid, dashed, and dot-dashed lines, respectively.}
\end{center}
\end{figure}

\section{Summary and Conclusions} \label{sec:conc}

We have presented a first comparison of high angular resolution observations of the 1.25~mm continuum emission from a solar active region (AR 12470) made by ALMA on 2015 December 18 with nearly simultaneous observations of the Mg II h doublet line obtained by IRIS. The two radiation regimes hold promise as powerful and complementary diagnostics of the structure and dynamics of the solar chromosphere. In comparing maps of the ALMA 1.25~mm  brightness temperature and the IRIS Mg II h2v radiation temperature we find a clear correlation between $T_B$(1.25~mm) and $T_{rad}$(Mg~II) although the Mg II radiation temperatures are offset to significantly lower values than the corresponding 1.25~mm brightness temperatures. Furthermore, the slope of the correlation is significantly smaller than unity. While the temperature offset is not understood, the slope $<1$ may, in part, be the result of the fact that the Mg II h line source function has a radiation scattering component, also compatible with the the significant scatter in the correlation. Masking the active region into pixels corresponding to sunspot umbra, ``quiet'' regions, and plage regions, we find that the correlation between $T_B$(1.25 mm) and $T_{rad}$(Mg II) is nonlinear. The slopes of the linear fits increase from 0.4, to 0.56, to 0.66 for sunspot, quiet, and plage pixels, respectively. Qualitatively, the larger slope measured in plage regions may be due to a higher degree of coupling of the source function to the gas in hotter, denser regions. These observations highlight the need for chromospheric models not only for quiet Sun conditions, but of active region conditions as well. Future observations will focus on expanding joint \mmsm and UV line observations to a variety of chromospheric environments and comparing them with the numerical models in detail as a means of better understanding the utility of the two wavelength regimes as complementary probes of the chromosphere.

\acknowledgments
The National Radio Astronomy Observatory is a facility of the National Science Foundation operated under cooperative agreement by Associated Universities, Inc. This paper makes use of the following ALMA data: ADS/JAO.ALMA \#2011.0.00020.SV. ALMA is a partnership of ESO (representing its member states), NSF (USA) and NINS (Japan), together with NRC (Canada), NSC and ASIAA (Taiwan), and KASI (Republic of Korea), in cooperation with the Republic of Chile. The Joint ALMA Observatory is operated by ESO, AUI/NRAO and NAOJ. IRIS is a NASA small explorer mission developed and operated by LMSAL with mission operations executed at NASA Ames Research center and major contributions to downlink communications funded by ESA and the Norwegian Space Centre. M.S. was supported by JSPS KAKENHI Grant Number JP17K05397. This work was partly carried out on the solar data analysis system and common-use data analysis computer system operated by ADC/ NAOJ. G.C. and B.D.P. are both supported by NASA contract NNG09FA40C (IRIS). M.L. acknowledges NSF grant AST-1312802 and NASA grant NNX14AK66G. 

\facilities{ALMA, IRIS, SDO}



\end{document}